\documentclass[a4paper]{jpconf}

\usepackage{graphicx}
\usepackage{citesort}

\begin{document}

\title{Sparticle masses from hadronic decays}

%\author{J M Butterworth$^1$, J R Ellis$^2$, A R Raklev$^3$}
\author{A R Raklev}

%\address{$^1$ Department of Physics and Astronomy, University College 
%London, Gower Street, London WC1E~6BT, UK}
%\address{$^2$ Theory Division, CERN, CH-1211 Geneva, Switzerland}
%\address{$^3$ Department of Physics and Technology, University of Bergen,
%All\'{e}gaten 55, N-5007 Bergen, Norway}
\address{Department of Physics and Technology, University of Bergen,
All\'{e}gaten 55, N-5007 Bergen, Norway}

\ead{Are.Raklev@ift.uib.no}

\begin{abstract}
We present our work on reconstructing sparticle masses in purely
hadronic decay chains, using the $k_T$ jet-algorithm on Monte Carlo
simulated events at LHC energies.
\end{abstract}

\section{Introduction}
The production and detection of supersymmetric particles is one of the
primary goals of the LHC experimental programs. As a proton-proton
collider we expect the dominant production at the LHC to be coloured
states: gluinos and squarks. Their decays through neutralinos and
charginos into leptonic final states have been widely
studied. However, much less work has been done on purely hadronic
decay chains, and decays which produce massive bosons such as $W$, $Z$
or a Higgs. Moreover, large branching ratios for neutralinos and
charginos into massive bosons have been shown to be a generic feature
of models that relax the GUT universality assumptions on Higgs masses
found in mSUGRA scenarios, and models that relax restrictions on
parameter space due to the measured Dark Matter density by having a
gravitino LSP~\cite{DeRoeck:2005bw}.

In~\cite{Butterworth:2007ke} we investigated the LHC potential for
finding supersymmetry, and for setting bounds on sparticle masses, in
decays of the type
\begin{equation}
\tilde q\to q\tilde\chi\to qB\tilde\chi_1^0,
\label{eq:chain}
\end{equation}
where $\tilde\chi$ is a neutralino/chargino intermediate and
$B=\{W,Z,h\}$. The difficulty with the hadronic decay of $B$ is the
large combinatorial background to the reconstruction - as a result of
the high jet activity expected at the LHC, a single event typically
has multiple boson candidates in the form of pairs of jets with the
correct invariant mass. This is in particular a challenge for the
measurement of sparticle masses through the invariant mass
distributions of the final states in decay chains such as
eq.~(\ref{eq:chain}). While a $Z$ can be reconstructed from decays
into electrons or muons at the expense of statistics, the missing
information from the neutrino in a leptonic $W$ decay was found to
make mass determination very difficult. The same is true for a low
mass Higgs decaying into taus.

To correctly identify massive bosons in decay chains such as
eq.~(\ref{eq:chain}) we have followed an idea put forward
in~\cite{Butterworth:2002tt} for analysing $WW$ scattering, using the
state-of-the-art $k_T$ jet-algorithm~\cite{Catani:1993hr} for
identifying a pair of collimated jets from the hadronic decay of a
boosted, massive particle. Following a short summary of the method
used, its successful application to a specific benchmark model and
decay chain, presented in detail in~\cite{Butterworth:2007ke}, will be
briefly discussed here.

\section{Method}

\subsection{Monte Carlo event generation}
In order to simulate sparticle production and top pair background at
the LHC, we use {\sc Pythia~6.408}~\cite{Sjostrand:2006za} with {\sc
CTEQ~5L} PDFs~\cite{Lai:1999wy} interfaced to the {\sc HZTool}
framework~\cite{hztool}, with some minor changes to allow for
simulations of SUSY scenarios. Decay widths and branching ratios for
the SUSY particles are calculated with {\sc
SDECAY~1.1a}~\cite{Muhlleitner:2003vg}.
The most important non-top Standard Model backgrounds consist of
multiple gauge boson and/or multiple jet production that are generated
using {\sc ALPGEN}~\cite{Mangano:2002ea}, {\sc
HERWIG~6.510}~\cite{Corcella:2002jc} and {\sc
Jimmy}~\cite{Butterworth:1996zw}.
%
%The parameters for the underlying event and the parton showers were
%those of the ATLAS tune of {\sc Pythia} and the CDF tune A of {\sc
%HERWIG} and {\sc Jimmy}, taken from~\cite{tunes}. These models have
%been shown to give a good description of a wide variety of data.

%In particular, the modelling of the internal jet structure by
%leading-logarithmic parton showers is known to be good for jets
%produced in $p\bar{p}$ collisions~\cite{tevjets}, $ep$ collisions and
%photo-production~\cite{herajets}, and in $e^+e^-$ annihilation events
%and $\gamma\gamma$ collisions~\cite{lepjets}.

\subsection{The $k_T$ jet-algorithm}
For each particle $k$ and pair of particles $(k,l)$, the $k_T$
jet-algorithm calculates the quantities
\begin{eqnarray}
d_{kB} & = & p_{Tk}^2, \nonumber \\
d_{lB} & = & p_{Tl}^2, \nonumber \\
d_{kl} & = & \min(p^2_{Tk},p^2_{Tl})R^2_{kl}/R^2,
\end{eqnarray}
where $p_{Tk}$ is the transverse momentum of particle $k$ with respect
to the beam axis and
\begin{equation}
R^2_{kl}=(\eta_k-\eta_l)^2+(\phi_k-\phi_l)^2.
\end{equation}
If $d_{kB}$ or $d_{lB}$ is the smallest, then particle $k$ or $l$ is
labelled a jet and removed from the list. If $d_{kl}$ is the smallest,
particles $k$ and $l$ are merged by adding their four-momenta. The
list is recalculated and the process is repeated until the list is
empty\footnote{The parameter $R$ plays a similar role to the
adjustable cone radius in cone algorithms. In our study we have used
$R=0.7$, which is a compromise between reconstruction efficiency for
the softest jets and mass resolution.}. The algorithm is infrared
safe, and has the additional benefit that each particle is uniquely
assigned to a single jet\footnote{Recently, a fast implementation of
the algorithm has been developed~\cite{Cacciari:2005hq}, which makes it
practical for use even in the very high multiplicity events expected
at the LHC.}.

\subsection{Signal isolation and jet identification}
To isolate the supersymmetry events from Standard Model background we
apply cuts on missing energy $\not\!\!E_T>300$~GeV, and require three
jets with $p_T>200,200,150$~GeV. To identify the massive boson in the
signal decay chain, we first require $p_T>200$~GeV for candidate jets
to ensure they are energetic enough to contain a boosted massive
particle. Then a cut is applied on the mass of the jet (calculated
from the four-vectors of the constituents) to ensure that it is in a
window around the nominal mass of the desired particle.

The final step is to decompose the jet into two sub-jets. The
information gained from this is the $y$ cut at which the sub-jets are
defined: $y \equiv d_{kl}/(p_T)^2$, where $p_T$ is the transverse
momentum of the candidate jet containing the sub-jets $k$ and $l$. In
the case of a genuine $W$, $Z$ or $h$ decay, the expectation for the
scale at which the jet is resolved into sub-jets (i.e., $yp_{T}^{2}$)
is ${\cal {O}}(M^{2})$, where $M$ is the boson mass. For QCD jets
initiated by a single quark or gluon, the scale of the splitting is
expected to be substantially below $p_T^2$, i.e., $y\ll 1$, since in
the region around the jet strongly-ordered QCD evolution dominates.

Although no detector simulation is employed in this analysis, the jet
mass and sub-jet scale cuts has previously been shown to be robust
against effects due to the calorimeter granularity and
resolution~\cite{allwood:thesis,stefanidis:thesis}.
 
\section{Results and conclusions}
\begin{figure}
\begin{center}
\includegraphics[width=16.7cm]{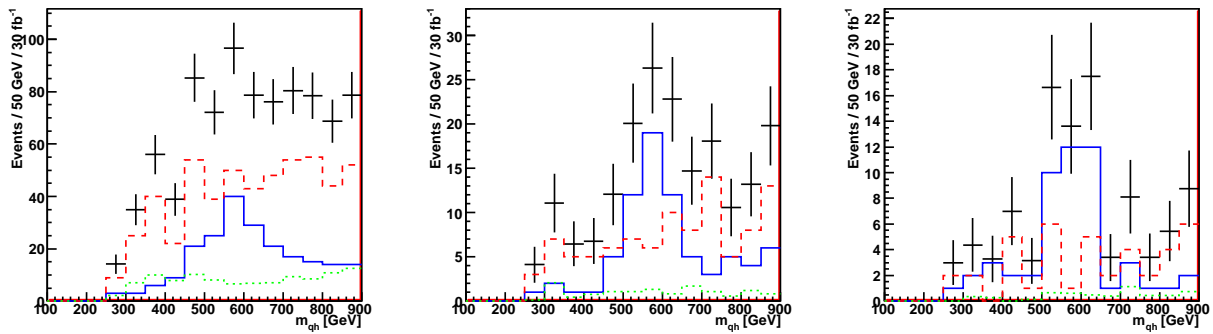}
\end{center}
\caption{\label{fig:cut_comp}
Invariant mass distributions for jet and Higgs candidate
combinations. Signal - blue, solid lines; SUSY background - red,
dashed lines; SM background - green, dotted lines.}
\end{figure}

As an example of the method, we present in
figure~\ref{fig:cut_comp} the invariant mass distribution of two jets
identified as coming from the squark and Higgs decays in the decay
chain (\ref{eq:chain}), for the SUSY benchmark scenario $\beta$, taken
from~\cite{DeRoeck:2005bw}. See~\cite{Butterworth:2007ke} for results
with a $W$ or $Z$ in the cascade and for other benchmarks. The squark
decay jet candidates are the jets in the event with $p_T>200$~GeV that
are not Higgs candidates, while the Higgs candidates are picked from
$b$-tagged\footnote{For details on the $b$-tagging procedure,
see~\cite{Butterworth:2007ke}} jets only (left), with an additional
jet mass cut of $110<m_j<140$~GeV (middle) and a sub-jet scale cut
$1.8<\log{(p_T\sqrt{y})}<2.1$ (right).

It is clear that the use of the jet mass and sub-jet scale cuts are
central in picking out the signal above the SUSY background. With the
application of the sub-jet scale cut, the expected edge in the
invariant mass distribution can be seen in the right panel, which
allows us to constrain the three sparticle masses involved in the
decay chain using standard edge analysis techniques.

We conclude that the sophistication of the $k_T$ jet-algorithm, giving
a jet mass with good resolution and the sub-jet scale of collimated
jets, can be central in reconstructing SUSY decay chains at the LHC
and in measuring or putting bounds on SUSY masses.

\ack
ARR acknowledges support from the European Community through a Marie
Curie Fellowship for Early Stage Researchers Training, and from the
Norwegian Research Council.

\section*{References}

\bibliographystyle{iopart-num}
\bibliography{BER}

\end{document}